\documentclass[twocolumn,english,prl]{revtex4}
\usepackage[T1]{fontenc}
\usepackage[latin9]{inputenc}
\usepackage{amsmath}
\usepackage{amssymb}
\usepackage{graphicx}
\usepackage{esint}

\makeatletter

\newcommand{\lyxdot}{.}

\makeatother

\usepackage{babel}
\begin{document}

\title{Transport blocking and topological phases using ac-magnetic fields}

\author{A. G\'{o}mez-Le\'{o}n}

\author{G. Platero}

\affiliation{Instituto de Ciencia de Materiales, CSIC, Cantoblanco, Madrid, 28049,
Spain.}

\date{\today}
\begin{abstract}
We analyze electron dynamics and topological properties of open double
quantum dots (DQDs) driven by circularly polarized ac-magnetic fields.
In particular we focus on the system symmetries which can be tuned
by the ac-magnetic field. Remarkably, we show that in the electron
spin resonance (ESR) configuration, where the magnetic fields in each
dot oscillate with a phase difference of $\pi$, charge localization
occurs giving rise to transport blocking at arbitrary intensities
of the ac field. The conditions for charge localization are obtained
by means of Floquet theory and related with quasienergies degeneracy.
We also demonstrate that a topological phase transition can be induced
in the adiabatic regime for a phase difference of $\pi$, either by
tuning the coupling between dots or by modifying the intensity of
the driving magnetic field. 
\end{abstract}
\maketitle

\section{Introduction}

The study of topological features in driven systems is a fascinating
topic, because of the emergence of non-trivial topologies once the
system interacts with external driving fields\citep{Inoue2010,Kitagawa2010,Lindner2010}.
Most of the topological features observed have been obtained under
the assumption of adiabatic evolution, but also other interesting
dynamical effects have been predicted in the non adiabatic regime
\citep{Grossmann1991,Gomez-Leon2011,PhysRevLett.105.086804,PhysRevB.82.205304}.
Among them, electron spin locking induced by linearly polarized ac
magnetic fields \citep{Gomez-Leon2011} or spin blockade induced by
ac magnetic fields\citep{PhysRevB.81.121306} have been recently proposed.
Therefore, investigation of different regimes in driven systems, and
the interplay between ac-fields, spatial symmetries and, topological
features in quantum systems becomes a very promising field which could
provide different mechanisms to manipulate electron charge and spin
in confined systems. Spin qubits can be produced in confined systems
as quantum dots, and their manipulation by driving the system with
ac fields has been one of the most recently active topics, both experimentally\citep{Koppens2006,Nowack30112007,Pioro2008,Perge2012,PhysRevLett.99.246601}
and theoretically\citep{PhysRevB.77.165312,PhysRevB.81.121306,PhysRevB.82.205304,Levitov,Loss,Nazarov}.

In this work we analyze the electron dynamics and current through
a double quantum dot (DQD) coupled to contacts and driven by circularly
polarized magnetic fields. We show that the driving field is able
to induce charge localization (CL) and then transport blocking for
arbitrary field intensity at electron spin resonance (ESR) configuration.
This localization effect depends critically on the symmetries of the
system and its appearance can also occur in larger size systems such
as arrays of quantum dots, linear ions traps\citep{Cirac95}, optical
lattices\citep{Struck2012}, and more generally, in tunnel coupled
systems with a pseudo-spin degree of freedom. We will show that in
our system, at resonance condition, CL occurs, if the ac magnetic
field oscillates in phase opposition in the different dots. We discuss
as well how CL is also reached off resonance for dots with different
Zeeman energies at a particular value of the ac frequency, i.e. at
the mean value of the Zeeman energy splittings of the dots. Coupling
the system to leads and applying a dc voltage gives rise to electronic
current through the DQD and then to characterize CL from the current.
We analyze the driven current by means of the Density Matrix (DM)
formalism within the framework of the Floquet theory. It allows to
effectively deal with the secular terms, and to characterize the effect
of decoherence in the current.

\begin{figure}
\includegraphics[scale=0.16]{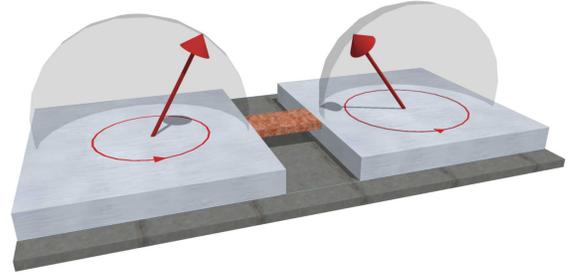}

\caption{\label{fig:Schematic}Schematic figure of a DQD coupled to circularly
polarized magnetic fields oscillating in phase opposition }
\end{figure}

In the last section we discuss how the topological properties of the
system are modified in the case where the phase of the ac magnetic
field within each dot differs in $\pi$. We demonstrate that a topological
phase transition can be induced in the adiabatic regime by tuning
the interdot coupling or by varying the intensity of the applied magnetic
field.

\section{Model:}

Our model describes a DQD in a strong Coulomb blockade regime, with
just up to one extra electron, coupled to leads. The states spanning
our DQD Hilbert space are: $\left\{ |\uparrow_{L}\rangle,\ |\downarrow_{L}\rangle,\ |\uparrow_{R}\rangle,\ |\downarrow_{R}\rangle,|0\rangle\right\} $.
The DQD is coupled to a magnetic field $\mathbf{B}^{i}=\left(B_{x}^{i}\left(t\right),B_{y}^{i}\left(t\right),B_{z}^{i}\right)$
by magnetic dipole interaction $g\mu_{B}\mathbf{S}\cdot\mathbf{B}\left(t\right)$
being $i=L,R$. The Zeeman splitting can be different for each dot
due to different g-factors, differences in the nuclei polarization
or inhomogeneities in the applied field. The ac-magnetic field, that
we assume circularly polarized, i.e. $B_{x}^{i}\left(t\right)=B_{ac,x}^{i}\cos\left(\omega t+\phi_{i}\right)$
and $B_{y}^{i}\left(t\right)=B_{ac,y}^{i}\sin\left(\omega t+\phi_{i}\right)$,
have a parameter $\phi_{i}$ characterizing the phase difference between
dots (Fig.\ref{fig:Schematic}).

The Hamiltonian reads: 
\begin{eqnarray}
H\left(t\right) & = & H_{S}\left(t\right)+H_{B}+V\label{eq:HamiltonianT}\\
H_{S}\left(t\right) & = & H_{0}+H_{ac}^{B}\left(t\right)+H_{dc}^{B}+H_{t_{LR}}\nonumber 
\end{eqnarray}
 being $H_{0}=\sum_{i,\sigma}\epsilon_{i}d_{\sigma,i}^{\dagger}d_{\sigma,i}$
the dot Hamiltonian for the $i$ dot, $H_{ac}^{B}\left(t\right)=\sum_{i,\mu}B_{ac,\mu}^{i}\left(t,\phi_{i}\right)S_{\mu,i}$,
$\left(\mu=x,y\right)$ the coupling of the electronic spin with the
external ac-field, $H_{dc}^{B}=\sum_{i}B_{z}^{i}S_{z,i}$ the Zeeman
splitting in the $i$ dot, and $H_{t_{LR}}=t_{LR}\sum_{\sigma,i\neq j}d_{\sigma,j}^{\dagger}d_{\sigma,i}$
the interdot tunneling Hamiltonian ( $\mu_{B}=g=\hbar=1$). The bath
Hamiltonian $H_{B}=\sum_{i,\sigma,K}\epsilon_{i,K}b_{i,K,\sigma}^{\dagger}b_{i,K,\sigma}$
represents fermionic reservoirs with operators $b_{i,K,\sigma}^{\dagger}/b_{i,K,\sigma}$
for electrons with $\epsilon_{K}$ energy and $K$ momentum, and $V=\sum_{i,\sigma,K}\Upsilon\left(b_{i,\sigma,K}^{\dagger}d_{i,\sigma}+h.c.\right)$
couples the DQD with the electron reservoirs. We consider Floquet
theory because the dynamics can be easily extracted from the quasienergy
spectrum, where degeneracies can be related with localization.

\paragraph{Symmetries:}

First we analyze the symmetries defined in the DQD, considering the
parity symmetry (PS) $\Pi:\left\{ x\rightarrow-x\right\} $ and the
generalized parity symmetry (GPS) $\Pi_{T}:\left\{ x\rightarrow-x,\ t\rightarrow t+T/2\right\} ,$
which usually play an important role in driven systems, and classify
the solutions according to a $\mathbb{Z}_{2}$ group\citep{Gomez-Leon2011}.
Applying the parity transformation to Eq.\ref{eq:HamiltonianT}, we
obtain the condition for parity invariance: $B_{z,L}=B_{z,R}$, $\epsilon_{L}=\epsilon_{_{R}}=0$,
$\phi=0$, and $B_{ac,\mu}^{L}=B_{ac,\mu}^{R}$. Note that if we consider
the generalized parity operation, an extra minus sign coming from
the time dependent term shows up, leading to a non invariant Hamiltonian.
The way to obtain $\Pi_{T}$ invariance is by considering $\phi=\phi_{2}-\phi_{1}=\pi$,
i.e. a $\pi$ difference of phase between the ac-fields in each dot.
This leads to a GP invariant Hamiltonian, but the PS is broken. Therefore
the difference of phase $\phi$ switches between $\Pi$ and $\Pi_{T}$
invariant systems.

In the present paper we shall consider $\Pi$ and $\Pi_{T}$ invariant
Hamiltonians, just by tuning $\phi$ to zero and $\pi$ respectively.
We also fix $B_{ac,x}^{L}=B_{ac,y}^{R}$, and just consider the possibility
of asymmetric Zeeman splittings (breaking both, PS and GPS).

Also an internal symmetry in a single dot can be defined. In the linearly
polarized case, a discrete $\mathbb{Z}_{2}$ symmetry is present due
to the time dependence of the field intensity $|\vec{B}\left(t\right)|$.
By contrary, in the circularly polarized case, the intensity $|\vec{B}|$
is constant and the system presents a continuous symmetry for rotations
along the $z$-axis $U_{R}\left(\theta\right)=e^{-i\theta S_{z}}$
. Furthermore, in this last case a time translation $t\rightarrow t^{\prime}$
is equivalent to a rotation along the z axis. This difference in the
internal symmetry of the single dot in the presence of magnetic fields
with different polarizations leads to the lack of dynamical spin locking\citep{Gomez-Leon2011}
in the circularly polarized case.

\paragraph*{Master equation for the reduced density matrix:}

When a quantum system is coupled to a dissipative bath such as a fermionic
reservoir, exchange of energy and information occurs, leading to decoherence
. In order to analyze how CL induced by circular ac-magnetic fields
is affected by decoherence, and how this is reflected in the current,
we develop a master equation within the framework of Floquet formalism,
that allows to analyze the current through the system. The influence
of phonons can be analyzed in a similar way, and an analogous effect
in the current is expected.

We start by considering the Liouville equation for the total DM in
the interaction picture. After some manipulations, the master equation
in the steady state for the reduced DM elements $\rho_{\alpha\beta}$
becomes (see details in \citep{PhysRevE.55.300,Hone2009} and Appendix\ref{sec:Appen-Master-equation}):
\begin{align}
i\varepsilon_{\alpha\beta}\rho{}_{\alpha\beta} & =\sum_{\mu\nu}\left(R_{\nu\beta,\mu\alpha}^{1}\left(0\right)+R_{\alpha\mu,\beta\nu}^{2}\left(0\right)\right)\rho{}_{\mu\nu}\nonumber \\
 & -R_{\alpha\mu,\nu\mu}^{1}\left(0\right)\rho{}_{\nu\beta}-R_{\nu\beta,\nu\mu}^{2}\left(0\right)\rho{}_{\alpha\mu},\label{eq:steady_state}
\end{align}

within the Markov approximation and Floquet basis, and being $R_{\alpha\beta,\mu\nu}^{1,2}\left(0\right)$
the transition rates due to the coupling with the contacts, which
couple the different Floquet states (see Eq.\ref{eq:Rates} in Appendix\ref{sec:Appen-Master-equation}).
In order to evaluate the current we calculate the time derivative
of the number of electrons $Q_{L}=\left(-e\right)\left(N_{L}+N_{dot,L}\right)$
in the left dot $N_{dot,L}=\sum_{\sigma}d_{\sigma,L}^{\dagger}d_{\sigma,L}$,
plus the left reservoir $N_{L}=\sum_{K,\sigma}b_{K,L,\sigma}^{\dagger}b_{K,L,\sigma}$.
Hence the average current is: $\langle I\rangle=\langle\dot{Q}_{L}\rangle=i\langle\left[H\left(t\right),Q_{L}\right]\rangle$.

\section{Results:}

We calculate the occupation probabilities in different states and
the current for different field configurations. The results are compared
with those of the closed system. We also consider different temperatures
and analyze its effect on the decoherence.

\begin{figure}[h]
 \includegraphics[scale=0.75]{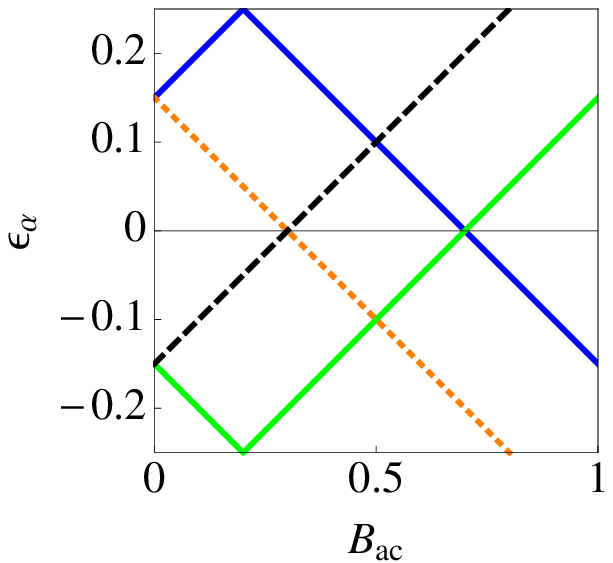}\includegraphics[scale=0.75]{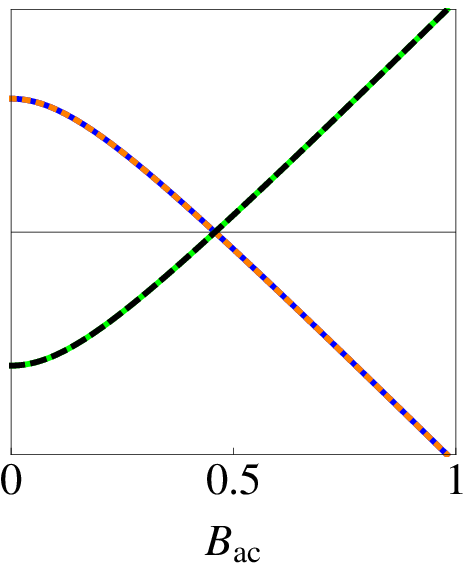}

\caption{\label{fig:Quasienergies1}(Color online) Quasienergies vs $B_{ac}$
for $\phi=0$ (left) and $\phi=\pi$ (right), in resonance condition.
The absence (presence) of $\Pi_{T}$ symmetry for $\phi=0$ ($\phi=\pi$),
leads to non-degenerate (doubly degenerate) quasienergies. The degeneracy,
present for all $B_{ac}$ in the right figure, drives the system to
CL. $t_{LR}=1/10$, and $B_{z,L}=B_{z,R}=\omega=1/2$.}
\end{figure}

\begin{figure}[h]
 \includegraphics[scale=0.75]{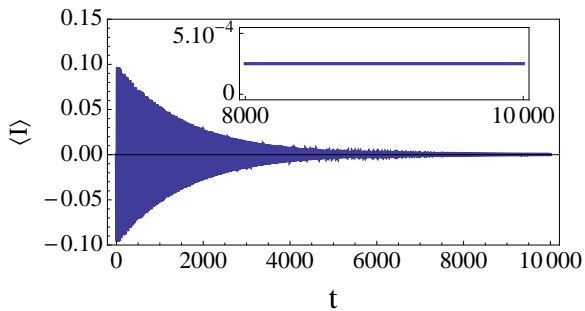}

\caption{\label{fig:Current1} Current vs $t$ for the case of $\Pi$ symmetry
($\phi=0$) at resonance condition. $B_{zL}=B_{zR}=\omega=0.5$, $B_{ac}=0.5$,
$t_{LR}=0.1$, $\Upsilon=0.01$, $T=10^{-3}$ and $\mu_{L}-\mu_{R}\gg T,B_{z},B_{ac}$.
The inset shows the current vs time in the steady state $\langle I\rangle\simeq2.5\times10^{-4}$
($3.5\times10^{-2}\text{pA}$). All values are in energy units of
$100\text{mT}$ ( $5.788\text{\ensuremath{\mu}eV}$). ($\omega\sim4.4\text{GHz}$),
the time unit scale is $\sim0.23\text{ns}$.}
\end{figure}

There are two cases to be considered, both in resonance condition
$B_{z,L}=B_{z,R}=\omega$. First, we deal with a $\Pi$ invariant
Hamiltonian ($\phi=0$). In this case, the quasienergies are not degenerate
for all $B_{ac}$ (see Fig.\ref{fig:Quasienergies1}, left), but just
isolated crossings appear for certain values of the field intensity.
The time evolution of the occupation probabilities in the closed system
(not shown in the paper) indicates that CL does not occur for this
field configuration independently of the field intensity and frequency,
and the electrons oscillate back and forth between the dots. Our numerical
calculation shows that all Floquet states become equally occupied
in the steady state. At a finite voltage $\mu_{L}-\mu_{R}\gg T,B_{z},B_{ac}$,
the current reaches a finite constant value (Fig.\ref{fig:Current1}),
which confirms the lack of CL.

\begin{figure}[h]
 \includegraphics[scale=0.75]{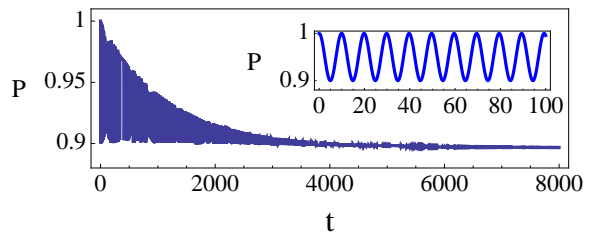}

\caption{\label{fig:Occupation_resonance}Occupation probability in the left
dot $P$ vs time for $B_{ac}=0.6$, in presence of decoherence due
to the leads. Note the spatial CL induced as a consequence of the
quasienergy degeneracy between Floquet states with opposite $\Pi_{T}$
symmetry. The inset shows the time evolution (Eq.\ref{eq:Localization})
for the closed system. $B_{z,L}=B_{z,R}=\omega=1/2$, $\Upsilon=0.01$,
$T=10^{-3}$, $t_{LR}=1/10$ and $\phi=\pi$, all in units of $100\text{mT}$
(i.e. $5.788\text{\ensuremath{\mu}eV}$) for $\mu_{L}-\mu_{R}\gg0$.}
\end{figure}

If we consider a $\Pi_{T}$ symmetry invariant Hamiltonian ($\phi=\pi$),
the quasienergies become doubly degenerated for all $B_{ac}$ (Fig.\ref{fig:Quasienergies1},
right). The occupation probabilities show that the system is in CL
regime (Fig.\ref{fig:Occupation_resonance}).

The existence of $\Pi_{T}$ symmetry allows the $\mathbb{Z}_{2}$
classification of the quasienergies, and then the double degeneracy
for all $B_{ac}$, driving the system to CL regime. The crossings
obtained between quasienergies with the same $\Pi$ or $\Pi_{T}$
symmetry are allowed by the Wigner-Von Neumann theorem because of
the continuous symmetry of a single dot in a circularly polarized
magnetic field, but they do not affect the occupation probabilities
of the Floquet states. The symmetry difference between this case,
and the case of a linearly polarized magnetic field (i.e. where the
single dot symmetry is discrete), is the reason for the absence of
dynamical spin locking in the former.

In the present case, i.e., for circularly driven fields we found a
very interesting result: CL is obtained for all values of $B_{ac}$
intensity. The reason is that the quasienergies manifold is degenerated
for all $B_{ac}$. Localization is improved by increasing the field
amplitude, as we will show below.

\begin{figure}[h]
 \includegraphics[scale=0.75]{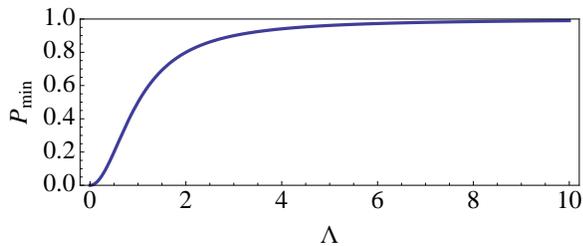}

\caption{\label{fig:Minimum-localization-expected} $P_{\text{min}}$ vs $\Lambda$
in presence of coupling with the leads for the steady state ($\phi=\pi$).}
\end{figure}

It can be seen by direct comparison of Fig.\ref{fig:Occupation_resonance}
and the inset, that the effect of decoherence due to the leads, is
to remove the coherent oscillations without breaking the localization
induced by the ac-field. In order to get a better insight of the effect
of decoherence on CL we obtain analytically the localization probability
for the closed system ($\phi=\pi$ and $B_{z,L}=B_{zR}=\omega$):
\begin{eqnarray}
P\left(t\right) & = & \left|\langle\uparrow_{L}|\Psi\left(t\right)\rangle\right|^{2}+\left|\langle\downarrow_{L}|\Psi\left(t\right)\rangle\right|^{2}\nonumber \\
 & = & 1+\frac{2t_{LR}^{2}\left(\cos\left(t\sqrt{B_{ac}^{2}+4t_{LR}^{2}}\right)-1\right)}{B_{ac}^{2}+4t_{LR}^{2}},\label{eq:Localization}
\end{eqnarray}
 Eq.\ref{eq:Localization} shows that CL depends on the ratio $B_{ac}/t_{LR}$.
From Fig.\ref{fig:Occupation_resonance} we assume that, in presence
of decoherence the localization is given by the minimum value of the
coherent oscillations in the closed system. We then can calculate
the expected CL for the open system (Fig.\ref{fig:Minimum-localization-expected}),
which is given by: $P_{\text{min}}=\text{min}\left(P\left(t\right)\right)=\Lambda^{2}/\left(1+\Lambda^{2}\right)$,
where $\Lambda=B_{ac}/\left(2t_{LR}\right)$.

From the last results, we conclude that the existence of $\Pi_{T}$
symmetry drives the system to charge localization, as $\omega$ is
in resonance with the Zeeman splitting. Once it is tuned off resonance,
the quasienergies split, and localization is destroyed.

Now we consider asymmetric Zeeman splittings. This is a very common
situation in real experiments, where Overhauser fields or different
g-factors make difficult to achieve a symmetric configuration. If
we calculate the Floquet spectrum for the Hamiltonian $H_{S}\left(t\right)$
(\ref{eq:HamiltonianT}), and look for the necessary condition for
degeneracy between quasienergies, we obtain: $\omega=\omega_{\text{Av}}:=\left(B_{z,L}+B_{z,R}\right)/2$.
Note that now $\Pi_{T}$ is always broken because of the asymmetric
Zeeman splittings. For the frequency condition $\omega=\omega_{\text{Av}}$,
the quasienergies overlap as in Fig.\ref{fig:Quasienergies1} (re-scaled
to the average Zeeman splitting), leading to CL regime at $\phi=\pi$.

\begin{figure}[h]
 \includegraphics[scale=0.75]{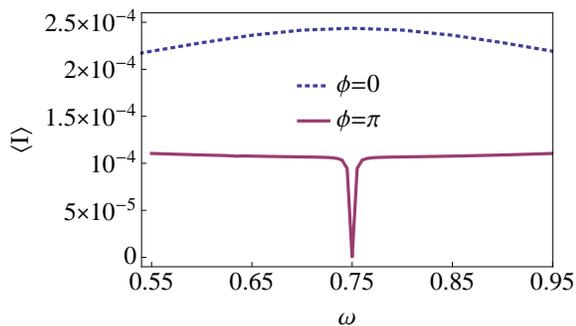}
\caption{\label{fig:Current-vs-frequency1}Current vs $\omega$ in the steady
state for asymmetric Zeeman splittings. $\langle I\rangle$ drops
to zero at $\omega=\omega_{\text{Av}}$ ($\omega_{\text{Av}}=0.75$)
for $\phi=\pi$, while for $\phi=0$ the current is finite. $B_{z,L}=0.5$,
$B_{z,R}=1$,$B_{ac}=0.55$, $t_{LR}=0.1$, and $\Upsilon=0.01$.
Different temperatures have been considered, but their effect just
affect to the transitory regime.}
\end{figure}

Finally Fig.\ref{fig:Current-vs-frequency1} shows $\langle I\rangle/\omega$
for $\phi=\left\{ 0,\pi\right\} $ and asymmetric Zeeman splittings.

This result demonstrates the critical difference in the current behavior
for $\phi=\left\{ 0,\pi\right\} $ at $\omega_{Av}$, where in the
last case,CL is induced and the current drops to zero. Now, we will
consider the adiabatic regime and we show that also the phase difference
is critical and determines the topology of the system.

\subsubsection{Topological phases in the adiabatic limit:}

In the adiabatic limit, spin systems coupled to magnetic fields show
a special feature. A non-trivial topological phase (characterized
by integer numbers $\mathbb{Z}$) arises for a single localized spin
whenever the adiabaticity parameter $\nu=\omega/2|\vec{B}|$ fulfills
$\nu<1$, being $|\vec{B}|=\sqrt{B_{z}^{2}+B_{ac}^{2}}$\citep{Bohm1}.
As the $\omega$ increases, the system is driven to the non-adiabatic
regime, inducing a topological phase transition to a region where
the topology is trivial.

In the present setup (for simplicity we consider $B_{z,L}=B_{z,R}$),
where the spin can oscillate between the two QDs, we characterize
the new topological features due to the tunnel coupling. Considering
both cases $\phi=\left\{ 0,\pi\right\} $, we calculate the first
Chern number $c_{1}$ that fully characterizes the topology of the
system, and the adiabaticity parameter $\nu^{\phi}$ (see appendix
for details). The result shows, in the $\phi=0$ case, that the tunneling
does not change the topological properties in the adiabatic limit
(same result as for a localized spin) with $c_{1}=\pm1$ for all values
of $t_{LR}$. In contrast, for the $\phi=\pi$ case, it is possible
to induce a topological phase transition within the adiabatic limit
just by tuning $\lambda=t_{LR}/|\vec{B}|$. This can be observed in
the first Chern number for each state: 
\begin{align*}
c_{1}\left(\lambda\right) & =\frac{1}{2}\left(1+\text{sign}\left(1-2\lambda\right)\right)\left(1,-1,1,-1\right)\\
 & =\begin{cases}
\left(1,-1,1,-1\right) & \forall\lambda<1/2\\
0 & \forall\lambda>1/2
\end{cases}
\end{align*}
 This result demonstrates that $t_{LR}=|\vec{B}|/2$ is a critical
point for the phase transition, and that within the adiabatic regime
we can induce a topological phase transition just by tuning the intensity
of the field.

In order to be in the adiabatic regime for the present configuration,
it is required to be out of resonance $\omega\ll B_{z}$. Therefore
the previous analysis concerning topology for $\phi=\left\{ 0,\pi\right\} $
applies for regimes where CL cannot occur. It shows, as we commented
previously, that different driving regimes in dynamical systems provide
different ways to manipulate electrons.

\section{Conclusions:}

We analyze the electron charge dynamics and the tunneling current
through a DQD attached to contacts and driven by a circularly polarized
magnetic field. We demonstrate both numerically and based in symmetry
arguments that tuning the phase difference between the ac magnetic
fields applied to the dots it is possible to achieve charge localization.
This effect is robust and perdure in presence of decoherence, induced
by the coupling with contacts. The effect of decoherence due to spin
baths will be analyzed in a further work\citep{Stamp}. Finally, we
demonstrate that out of resonance, in the adiabatic limit, we can
drive our system to different topological phases by tuning the ratio
$t_{LR}/|\vec{B}|$.

In summary, two level systems such as QDs coupled by tunneling and
interacting with ac-magnetic fields show interesting features which
depend drastically on the field polarization. The nature of the magnetic
dipolar coupling $-\vec{\mu}\cdot\vec{B}$ and the interplay between
spatial and spin degrees of freedom coupled through the ac-magnetic
field allow to achieve charge localization for arbitrary intensity
of the ac magnetic field in the ESR regime. Also, in the adiabatic
limit, we show that a topological phase transition can be induced.
The present results demonstrate the huge horizon of possibilities
obtained by combining ac magnetic fields and QD arrays, with direct
application in quantum computation with spin qubits and also in topological
quantum computation, although the latter should be studied further
for the case of non-abelian gauge theories. 
\begin{acknowledgments}
We thank R. Sanchez, S. Kohler and C. E. Creffield for critical reading
of the manuscript. We acknowledge MAT 2011-24331 and ITN, grant 234970
(EU) for financial support. A. G\'{o}mez-Le\'{o}n acknowledges JAE program. 
\end{acknowledgments}
\appendix
%dummy comment inserted by tex2lyx to ensure that this paragraph is not empty

\section{\label{sec:Appen-Master-equation}Master equation. Detailed calculation}

Following previous works \citep{PhysRevE.55.300,Hone2009}, we consider
the Liouville equation for the density matrix and weak coupling with
the leads. We calculate the reduced density matrix by tracing out
the bath degrees of freedom. The assumptions at this point are the
approximate factorization of the density matrix between bath and system,
and Markov approximation. The density matrix of the bath is given
by a fermionic thermodynamical ensamble in equilibrium.

We assume that our interaction Hamiltonian is linear in the system
and lead operators: 
\begin{equation}
V\equiv\sum_{j}F_{j}Q_{j}=\Upsilon\sum_{K,i,\sigma}\left(b_{K,i,\sigma}^{\dagger}d_{i,\sigma}-b_{i,\sigma,K}d_{i,\sigma}^{\dagger}\right)\label{eq:interaction_op-1}
\end{equation}
 being $\Upsilon$ the coupling to the leads, $F_{j,K}$ a lead operator,
and $Q_{j}$ a system operator ($K$ is the momentum of the electron
in the lead). We have considered multindex notation, being $\pm j=\left(K,i,\sigma,\pm\xi\right)$,
$i=L,R$, $\sigma=\uparrow,\downarrow$, and $\xi=\pm$ (this means
annihilation/creation operator for the system). Identifying the coefficients
we obtain $Q_{i,\sigma,+}\equiv d_{i,\sigma}$, $Q_{i,\sigma,-}\equiv-d_{i,\sigma}^{\dagger}$,
$F_{K,i,\sigma,+}\equiv b_{K,i,\sigma}^{\dagger}$, and $F_{K,i,\sigma,-}\equiv b_{K,i,\sigma}$.

When we trace out the bath degrees of freedom, the terms involving
the statistical ensambles can be written as: 
\begin{align}
G_{1}\left(t-\tau,t\right) & =\sum_{j,l}Tr_{B}\left\{ F_{j}\tilde{F}_{l}\left(t-\tau,t\right)\rho_{B}\right\} \label{eq:Thermal}\\
G_{2}\left(t-\tau,t\right) & =\sum_{j,l}Tr_{B}\left\{ \tilde{F}_{l}\left(t-\tau,t\right)F_{j}\rho_{B}\right\} ,\nonumber 
\end{align}
 The calculation of these terms, also integrating in $\tau$, leads
to: 
\begin{align*}
g_{1}\left(m\omega-\varepsilon_{\mu\nu}\right)_{+,i} & =\pi D_{i}\left(1-n_{F}\left(m\omega-\varepsilon_{\mu\nu}-\mu_{i}\right)\right)\\
 & =g_{2}\left(m\omega-\varepsilon_{\mu\nu}\right)_{-,i}\\
g_{2}\left(m\omega-\varepsilon_{\mu\nu}\right)_{+,i} & =\pi D_{i}n_{F}\left(m\omega-\varepsilon_{\mu\nu}-\mu_{i}\right)\\
 & =g_{1}\left(m\omega-\varepsilon_{\mu\nu}\right)_{-,i}
\end{align*}
 being $D_{i}$ the density of states in the $i$ lead, that we assumed
to be constant in our regime, $\mu_{i}$ the chemical potential, and
$n_{F}\left(E\right)=\left(1+e^{E/\left(K_{B}T\right)}\right)^{-1}$
the usual Fermi distribution function.

The rates of the master equation require the calculation of the matrix
elements $\langle\phi_{\alpha}\left(t\right)|Q_{j}\tilde{Q}_{l}\left(t-\tau,t\right)\rho_{S}\left(t\right)|\phi_{\beta}\left(t\right)\rangle$,
being $|\phi_{\alpha}\left(t\right)\rangle$ the Floquet states of
the closed system. Using Fourier series due to the periodicity of
Floquet states, and defining $Q\left(n\right)_{j,\alpha\beta}\equiv\frac{1}{T}\int_{0}^{T}dte^{in\omega t}\langle\phi_{\alpha}\left(t\right)|Q_{j}|\phi_{\beta}\left(t\right)\rangle$.
We obtain the full master equation in Floquet basis: 
\begin{eqnarray}
\left(\partial_{t}+i\varepsilon_{\alpha\beta}\right)\rho\left(t\right)_{\alpha\beta} & = & \sum_{\mu\nu}\left(R_{\nu\beta,\mu\alpha}^{1}\left(t\right)+R_{\alpha\mu,\beta\nu}^{2}\left(t\right)\right)\rho\left(t\right)_{\mu\nu}\label{eq:Master_eq_expanded}\\
 &  & -R_{\alpha\mu,\nu\mu}^{1}\left(t\right)\rho\left(t\right)_{\nu\beta}-R_{\nu\beta,\nu\mu}^{2}\left(t\right)\rho\left(t\right)_{\alpha\mu},\nonumber 
\end{eqnarray}
 where $R_{\alpha\beta,\mu\nu}^{1,2}\left(t\right)=\sum_{M}R_{\alpha\beta,\mu\nu}^{1,2}\left(M\right)e^{-i\omega Mt}$,
and the Fourier coefficients are: 
\begin{eqnarray}
R_{\alpha\beta,\mu\nu}^{1}\left(M\right) & = & \left|\Upsilon\right|^{2}\sum_{i,\sigma,\xi}\sum_{m}Q\left(M+m\right)_{i\sigma\xi,\alpha\beta}Q\left(m\right)_{i\sigma\xi,\mu\nu}^{*}\label{eq:Rates}\\
 &  & \times g_{1}\left(\varepsilon_{\mu\nu}-m\omega\right)_{\xi,i}\nonumber \\
R_{\alpha\beta,\mu\nu}^{2}\left(M\right) & = & \left|\Upsilon\right|^{2}\sum_{i,\sigma,\xi}\sum_{m}Q\left(M+m\right)_{i\sigma\xi,\alpha\beta}Q\left(m\right)_{i\sigma\xi,\mu\nu}^{*}\nonumber \\
 &  & \times g_{2}\left(\varepsilon_{\mu\nu}-m\omega\right)_{\xi,i}\nonumber 
\end{eqnarray}
 The approximation considered in the paper $R_{\alpha\beta,\mu\nu}^{1,2}\left(M\right)=R_{\alpha\beta,\mu\nu}^{1,2}\left(0\right)$,
can be assumed if the quasienergies are far from the boundaries of
the Brillouin zone.

\section{\label{sec:Adiabatic-limit}Adiabatic limit}

We define the adiabatic parameter following \citep{Mustafazadeh97}
as: 
\begin{eqnarray*}
\omega_{c}: & = & \sup\left\{ |A_{mn}\left(t\right)|:\ n\neq m=1,...,4\right\} ,\\
 &  & t\in\left[\tau_{1},\tau_{2}\right]\subseteq\left[0,2\pi/\omega\right]\\
A_{mn} & = & \langle m;t|\frac{d}{dt}|n;t\rangle\\
\nu & := & \frac{\omega_{c}}{\omega_{0}}
\end{eqnarray*}
 being $\omega_{0}$ the level spacing to the first excited level
and $\tau$ the total time needed for a cycle. In the case $\phi=0\rightarrow\omega_{c}^{\phi=0}=\Omega/2$
and the adiabaticity parameter is $\nu^{\phi=0}=\Omega/2t_{LR}$ (taking
$\omega_{0}^{\phi=0}=t_{LR}$). The case $\phi=\pi$ is very different
because the level splitting to the first excited state is no longer
given by the tunneling parameter. Hence, 
\begin{eqnarray*}
\omega_{0}^{\phi=\pi} & = & \frac{1}{2}\sqrt{|\vec{B}|^{2}+4t_{LR}^{2}+4t_{LR}B_{z}}\\
 &  & -\frac{1}{2}\sqrt{|\vec{B}|^{2}+4t_{LR}^{2}-4t_{LR}B_{z}},
\end{eqnarray*}
 
\[
\omega_{c}^{\phi=\pi}=\omega|\vec{B}|/\left(2\sqrt{|\vec{B}|^{2}+4t_{LR}^{2}}\right),
\]
 and 
\[
\nu^{\phi=\pi}=\frac{\omega_{c}^{\phi=\pi}}{\omega_{0}^{\phi=\pi}}
\]
 The adiabatic regime requires $\nu^{\phi}\ll1$, and then small frequencies
compared with all the other energy scales. For the adiabaticity condition
we have assumed $t_{LR}<B_{z}$.

In conclusion we state that for frequencies small enough (far from
resonance), varying the tunneling parameter, or the intensity of the
magnetic field, a topological phase transition can be induced.

 \bibliographystyle{phaip}

\end{document}